\documentclass[aps,prl,twocolumn,superscriptaddress]{revtex4}
\usepackage{graphicx}

\def\be{\begin{equation}}       \def\ee{\end{equation}}
\def\bea{\begin{eqnarray}}      \def\eea{\end{eqnarray}}

\begin{document}
\title{Spin Waves and magnetic exchange interactions in insulating Rb$_{0.89}$Fe$_{1.58}$Se$_2$}
\author{Miaoyin Wang}
\affiliation{Department of Physics and Astronomy, The University of Tennessee, Knoxville,
Tennessee 37996-1200, USA}
\author{Chen Fang}
\affiliation{Department of Physics, Purdue University, West Lafayette, Indiana 47907, USA}
\author{Dao-Xin Yao}
\affiliation{State Key Laboratory of Optoelectronic Materials and Technologies,
Sun Yat-sen University, Guangzhou 510275, China}
\author{GuoTai Tan}
\affiliation{Department of Physics and Astronomy, The University of Tennessee, Knoxville,
Tennessee 37996-1200, USA}
\affiliation{College of Nuclear Science and Technology, Beijing Normal University, Beijing 100875, China}
\author{Leland W. Harriger}
\affiliation{Department of Physics and Astronomy, The University of Tennessee, Knoxville,
Tennessee 37996-1200, USA}
\author{Yu Song}
\affiliation{Department of Physics and Astronomy, The University of Tennessee, Knoxville,
Tennessee 37996-1200, USA}
\author{Tucker Netherton}
\affiliation{Department of Physics and Astronomy, The University of Tennessee, Knoxville,
Tennessee 37996-1200, USA}
\author{Chenglin Zhang}
\affiliation{Department of Physics and Astronomy, The University of Tennessee, Knoxville,
Tennessee 37996-1200, USA}
\author{Meng Wang}
\affiliation{Beijing National Laboratory for Condensed Matter Physics and Institute of
Physics, Chinese Academy of Sciences, P. O. Box 603, Beijing 100190, China}
\affiliation{Department of Physics and Astronomy, The University of Tennessee, Knoxville,
Tennessee 37996-1200, USA}
\author{Matthew B. Stone}
\affiliation{Neutron Scattering Science Division, Oak Ridge National Laboratory, Oak
Ridge, Tennessee 37831-6393, USA}
\author{Wei Tian}
\affiliation{Ames Laboratory and Department of Physics and Astronomy
            Iowa State University, Ames, Iowa 50011, USA}
\author{Jiangping Hu}
\affiliation{Department of Physics, Purdue University, West Lafayette, Indiana 47907, USA}
\affiliation{Beijing National Laboratory for Condensed Matter Physics and Institute of
Physics, Chinese Academy of Sciences, P. O. Box 603, Beijing 100190, China}
\author{Pengcheng Dai}
\email{pdai@utk.edu}
\affiliation{Department of Physics and Astronomy, The University of Tennessee, Knoxville,
Tennessee 37996-1200, USA}
\affiliation{Neutron Scattering Science Division, Oak Ridge National Laboratory, Oak
Ridge, Tennessee 37831-6393, USA}
\affiliation{Beijing National Laboratory for Condensed Matter Physics and Institute of
Physics, Chinese Academy of Sciences, P. O. Box 603, Beijing 100190, China}


\maketitle
{\bf The discovery of alkaline iron selenide
$A$Fe$_{1.6+x}$Se$_2$ ($A=$ K, Rb, Cs) superconductors \cite{jgguo,krzton,mhfang,afwang,gfcheng2011} has generated considerable excitement in the condensed matter physics community because superconductivity in these materials may have a different origin
from the sign reversed $s$-wave electron pairing mechanism \cite{yzhang,tqian,Dxmou},
a leading candidate proposed for all other Fe-based superconductors \cite{mazin11,mazin2011n}.
Although $A$Fe$_{1.6+x}$Se$_2$ are isostructural with the metallic antiferromagnetic (AF)
iron pnictides such as (Ba,Ca,Sr)Fe$_2$As$_2$ \cite{johnston,cruz},
they are insulators near $x=0$ \cite{mhfang,afwang,gfcheng2011}
and form a $\sqrt{5}\times\sqrt{5}$ blocked AF structure (Fig. 1a) completely different from the
iron pnictides \cite{haggstrom,bacsa,wbao1,pomjakushin1,wbao2}.
If magnetism is responsible for superconductivity of all
iron-based materials \cite{mazin2011n}, it is important to determine their common magnetic features.
Here we use neutron scattering to map out spin waves in the AF
insulating Rb$_{0.89}$Fe$_{1.58}$Se$_2$. We find that although Rb$_{0.89}$Fe$_{1.58}$Se$_2$
has a N$\rm \acute{e}$el temperature ($T_N=475$ K) much higher than that of the iron pnictides
($T_N\leq 220$ K), spin waves for both classes of materials have similar zone boundary energies \cite{lharriger,jzhao,raewings}.
A comparison of the fitted effective exchange couplings using
a local moment Heisenberg Hamiltonian
 in Rb$_{0.89}$Fe$_{1.58}$Se$_2$, (Ba,Ca,Sr)Fe$_2$As$_2$  \cite{lharriger,jzhao,raewings}, and iron chalcogenide Fe$_{1.05}$Te \cite{lipscombe} reveals that their next nearest neighbor (NNN) exchange couplings are similar.  Therefore, superconductivity in
all Fe-based materials may have a common magnetic origin that is intimately associated with the
 NNN magnetic exchange interactions,
even though they have metallic or insulating ground states, different AF orders and electronic band structures.
}

Soon after the discovery of superconductivity in iron pnictides \cite{kamihara}, calculations and experiments have found that electronic band structures of these materials are composed of hole and electron Fermi pockets near $\Gamma(0,0)$ and $M(1,0)/M(0,1)$ points, respectively \cite{mazin2011n}.  As a consequence,
sign reversed quasiparticle excitations between the hole and electron pockets
can induce $s^{\pm}$-wave superconductivity, giving rise to
a neutron spin resonance at the in-plane wave vector $Q = (1,0)$ (Fig. 1c) \cite{maier,korshunov,christianson}. If sign reversed
electron-hole pocket excitations between $\Gamma(0,0)$ and $M(1,0)/M(0,1)$ points are necessary for superconductivity,
superconductivity in alkaline iron selenides
should have a different microscopic origin since angle resolved photoemission experiments measurements
on these materials reveal only electron Fermi
surfaces at $M(1,0)/(0,1)$ points and no hole Fermi pockets
at $\Gamma(0,0)$ point \cite{yzhang,tqian,Dxmou}.  On the other hand, if AF spin excitations are
responsible for superconductivity in Fe-based superconductors \cite{mazin2011n,seo08},
one would expect that spin waves in
the parent compounds of different classes of Fe-based superconductors have a common magnetic
origin associated with superconductivity.  Previous work on spin waves of
(Ba,Ca,Sr)Fe$_2$As$_2$  \cite{lharriger,jzhao,raewings} and Fe$_{1.05}$Te \cite{lipscombe} suggests that the
NNN exchange couplings in these materials are similar.  Since the insulating $A$Fe$_{1.6+x}$Se$_2$ has completely different
magnetic structure  and static ordered moment (Fig. 1) from those of (Ba,Ca,Sr)Fe$_2$As$_2$ and Fe$_{1.05}$Te \cite{johnston}, it is important to determine if
its effective magnetic exchange couplings are similar to these materials.

Here we report inelastic neutron scattering studies of spin waves in the insulating Rb$_{0.89}$Fe$_{1.58}$Se$_2$
with $T_N=475$ K.  Our neutron diffraction
measurements on the sample confirmed the previously proposed
Fe$_4$ block AF checkerboard structure (Fig. 1a) \cite{wbao1}.
Since the ferromagnetic (FM) Fe$_4$ block in the $\sqrt{5}\times\sqrt{5}$ superlattice unit cell
can have either left or right chirality (Figs. 1a and 1b), one expects to observe four AF Bragg peaks stemming from
each of the chiralities.  Figure 1c shows the expected AF peaks from the left chirality in reciprocal space using
the orthorhombic unit cell similar to that of iron pnictides \cite{lharriger,jzhao,raewings}, where they occur at
$(H_o,K_o,L_o)=(0.2+m,0.6+n,L_o); (-0.2+m,-0.6+n;L_o);(0.6+m,-0.2+n,L_o);(-0.6+m,0.2+n,L_o$ ($m,n=\pm 2,\pm 4, \cdots$, and $L_o=\pm 1,\pm 3, \cdots$).
Considering both chiralities for the AF order, there are eight
Bragg peaks at wave vectors $(H_o,K_o,L_o)=(\pm 0.2+m,\pm 0.6+n,L_o)$ and
$(H_o,K_o,L_o)=(\pm 0.6+m,\pm 0.2+n,L_o)$ from the block AF checkerboard structure (Fig. 1d), where
the odd values of $L_o$ indicate AF coupling along the $c$-axis direction \cite{wbao1,pomjakushin1,wbao2}. Therefore,
acoustic spin waves in the AF ordered phase of Rb$_{0.89}$Fe$_{1.58}$Se$_2$ should stem from these eight Bragg peaks.

\begin{figure}[t]
\includegraphics[scale=.4]{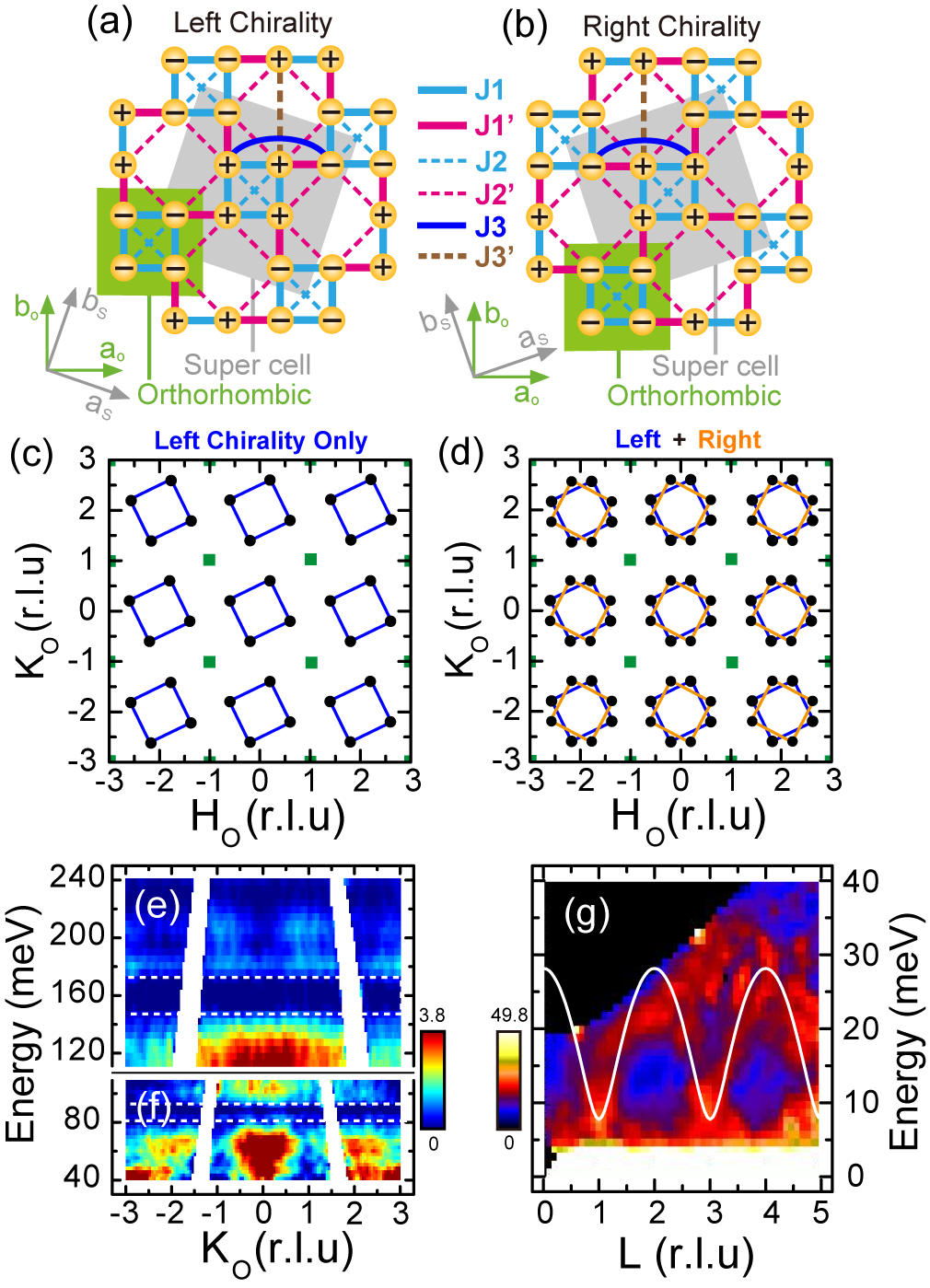}
\caption{ The AF spin structure, reciprocal space, and $c$-axis spin waves of the insulating Rb$_{0.89}$Fe$_{1.58}$Se$_2$.
Our neutron scattering experiments were carried out on the ARCS
chopper spectrometer at the Spallation Neutron Source, Oak Ridge National Laboratory. We
co-aligned 2.7 g of single crystals grown by self-flux (with mosaic of $\sim$6$^\circ$).
The incident beam energies were $E_i = 80, 140, 250, 440$ meV, and mostly with $E_i$ parallel to
the $c$-axis. Spin-wave intensities were normalized
to absolute units using a vanadium standard (with 30\% error).
We define the wave vector Q at $(q_x, q_y, q_z)$ as $(H_o;K_o;L_o)=(q_xa_o/2\pi;q_ya_o/2\pi;q_zc_o/2\pi)$ rlu,
where $a_o=5.65$ and $c_o=14.46$ \AA\ are the orthorhombic cell lattice parameters.
The AF spin structures are shown for (a) left and (b) right chirality.
The $\sqrt{5}\times\sqrt{5}$ superlattice structure is marked as grey with lattice parameter $a_s=8.933$ \AA.
The orthorhombic lattice cell is shaded green.
The effective NN, NNN, NNNN exchange couplings are marked as $J_1/J_1^\prime$, $J_2/J_2^\prime$, and $J_3/J_3^\prime$, respectively.
(c) The $[H_o,K_o]$ reciprocal space with the expected AF Bragg peaks from the left chirality.
The green squares show nuclear Bragg peak positions.
(d) Expected Bragg peaks for both chiralities.
(e,f) Spin waves projected onto the $K_o$-$E$ plane with $H_o$ integration from -2 to -1.
The scattering were measured with $E_i=440, 250$ meV, respectively.
(g) $c$-axis spin-wave dispersion projected on the $L$-$E$ plane with $H_o$ integration from 0.5 to 0.7 and $K$ integration from
0 to 0.4.  The solid line is the calculated $c$-axis dispersion using effective exchange couplings discussed in the main text.
}
\end{figure}

\begin{figure}[t]
\includegraphics[scale=.48]{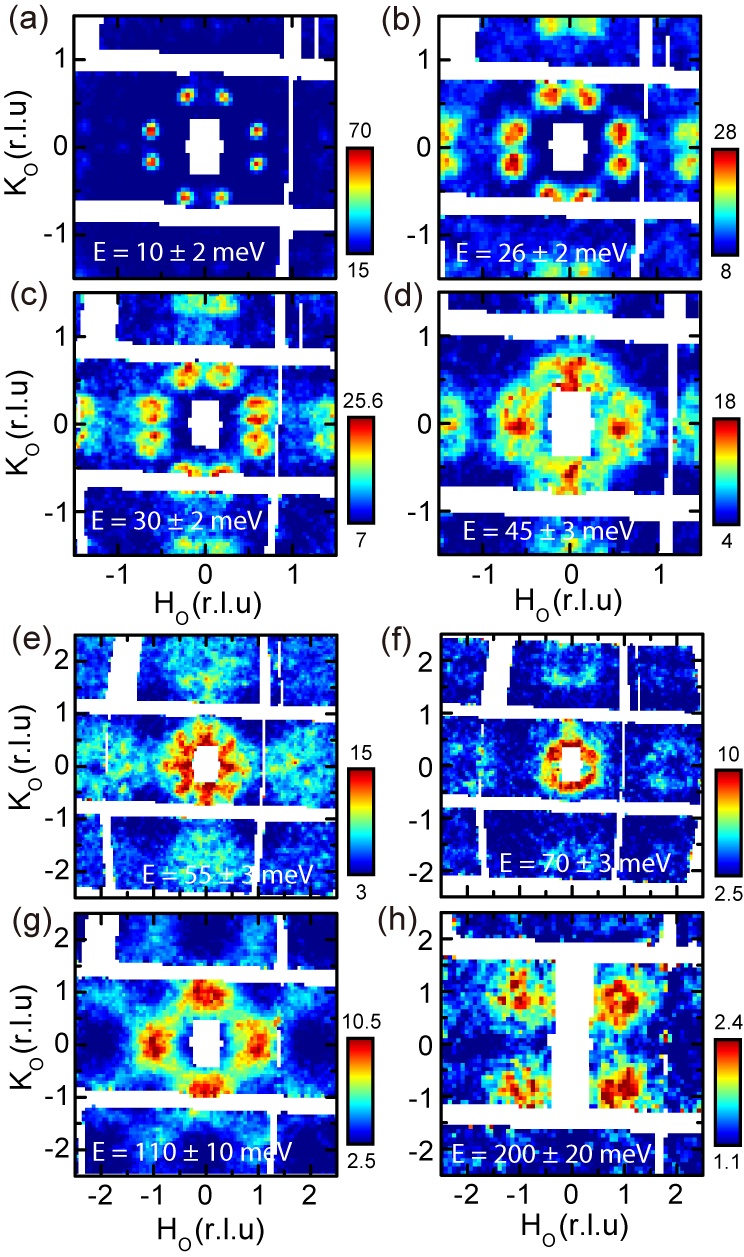}
\caption{ Wave-vector dependence of spin-wave excitations at different energies in the $[H_o,K_o]$ scattering plane
for Rb$_{0.89}$Fe$_{1.58}$Se$_2$ at 10 K.  Spin wave excitations at energies
(a) $E=10\pm2$; (b) $E=26\pm2$; (c) $E=30\pm2$; (d) $E=45\pm3$; (e) $E=55\pm3$; (f) $E=70\pm3$;
(g) $E=110\pm10$; and (h) $E=200\pm20$ meV.
(a)-(c), (d)-(f),(g),(h) were obtained with
$E_i=80$, 140, 250, and 440 meV, respectively, along the $c$-axis.
The vertical color bars indicate intensity scale in mbarns/sr/meV/f.u.
}
\end{figure}

Before mapping out the wave vector dependence of spin waves in Rb$_{0.89}$Fe$_{1.58}$Se$_2$, we first determine their overall energy bandwidth  and the effective $c$-axis coupling.  Figures 1e and 1f show the background subtracted scattering
projected in the wave vector ($Q = [-1.5,K_o]$) and energy plane.
One can see three clear plumes of scattering arising from the in-plane AF zone centers $Q = (0,-2), (0,0),$ and (0,2) rlu.
With increasing energy, spin waves are gapped at energies between 75 and 95 meV (Fig. 1f) and between 150 and 170 meV (Fig. 1e).
The zone boundary spin wave energies are around 220 meV (Fig. 1e). Therefore,
in spite of the large differences in  N$\rm \acute{e}$el temperatures of Rb$_{0.76}$Fe$_{1.6}$Se$_2$ ($T_N=475$ K) \cite{wbao1,pomjakushin1,wbao2}, (Ba,Ca,Sr)Fe$_2$As$_2$ ($T_N\leq 220$ K) \cite{lharriger,jzhao,raewings}, and Fe$_{1.05}$Te ($T_N\approx 70$ K) \cite{lipscombe}, their zone boundary spin wave energies are rather similar.
To estimate the AF coupling strength along the $c$-axis, we show in Fig. 1g spin waves projected in the wave vector
$Q = [0.6,0.2,L_o]$ and energy space. One can see clear dispersive spin waves stemming from AF positions $L_o=1,3,5$ that reach
the zone boundary energy near 30 meV.

\begin{figure}[t]
\includegraphics[scale=.48]{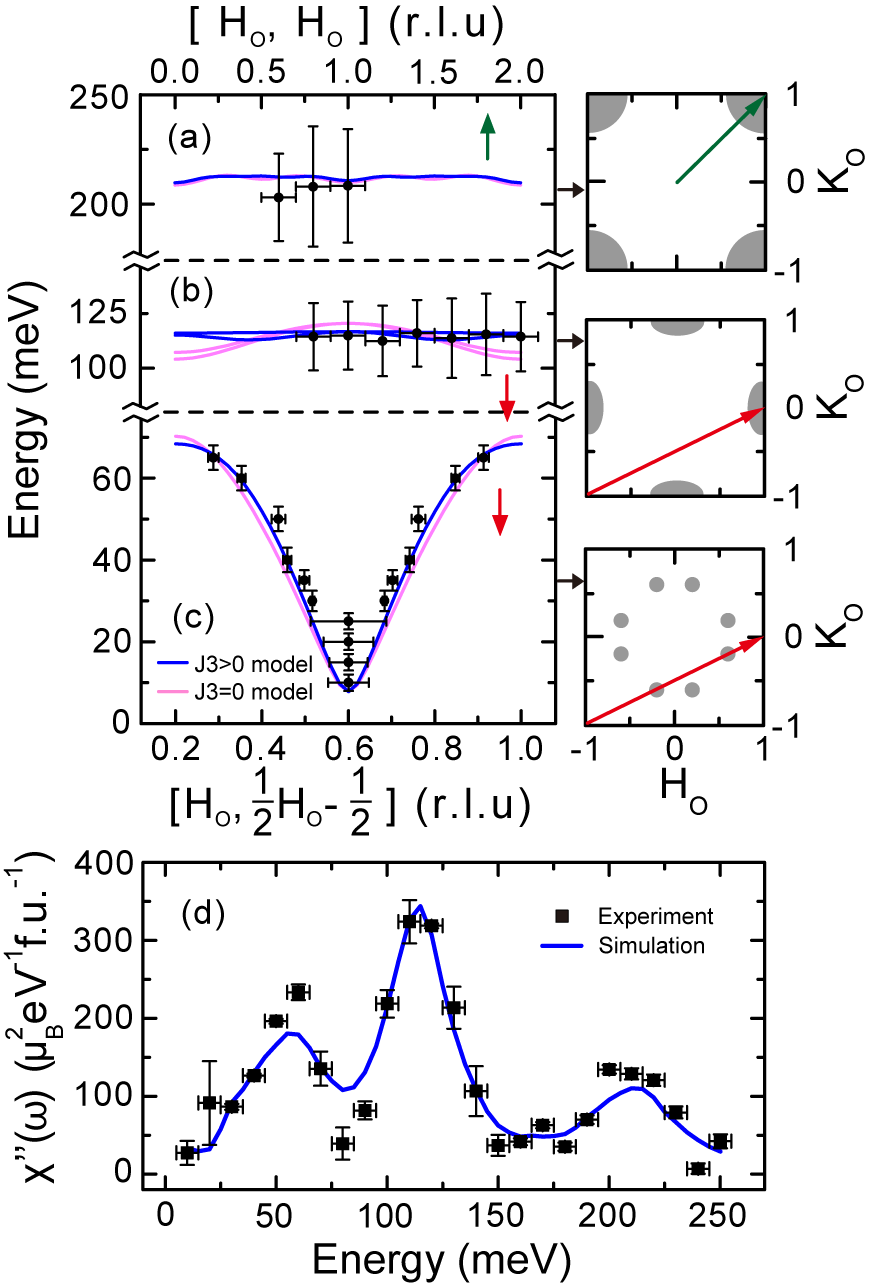}
\caption{ Spin-wave dispersions of Rb$_{0.89}$Fe$_{1.58}$Se$_2$ and fits using Heisenberg Hamiltonian.
Spin-wave dispersions obtained by cutting along high-symmetry directions marked in the right panels for
(a) highest energy optical energy band; (b) medium energy optical energy band; and (c) acoustic spin wave mode.
The blue solid lines show fits with $J_3>0$, while the pink solid lines are fits with $J_3=0$. (d) The energy dependence of the
local susceptibility and our model calculation of the local susceptibility.
}
\end{figure}

\begin{figure}[t]
\includegraphics[scale=.48]{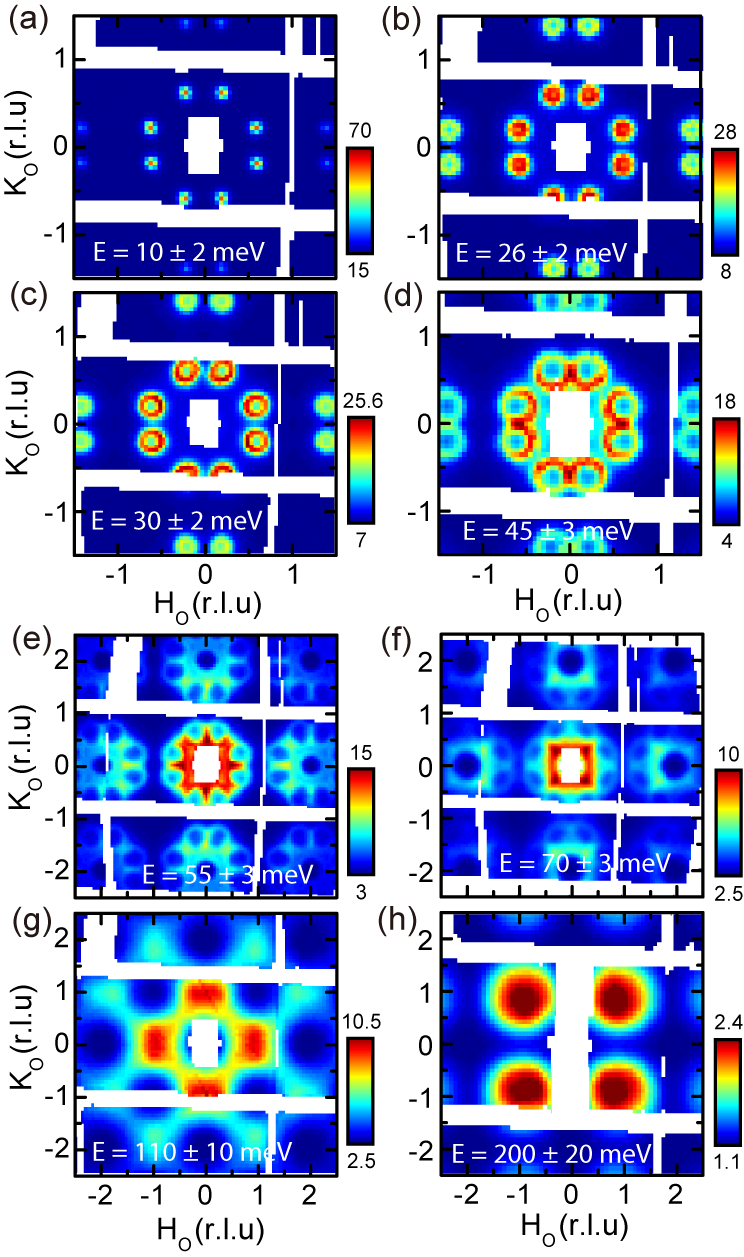}
\caption{ Calculated wave-vector dependence of the spin waves in the $[H_o,K_o]$ scattering for identical energies
as that of Fig. 2.  The instrumental resolution is convoluted with the Heisenberg Hamiltonian.
}
\end{figure}

To see the evolution of spin waves with increasing energy, we show in Fig. 2
the two-dimensional constant-energy ($E$) images of spin waves in the $[H_o,K_o]$ plane
for various incident beam energies ($E_i$).
From their $c$-axis dispersion (Fig. 1g), we know that
spin waves in Rb$_{0.89}$Fe$_{1.58}$Se$_2$ are three-dimensional
similar to that in (Ba,Ca,Sr)Fe$_2$As$_2$ \cite{lharriger,jzhao,raewings}
and center at AF wave vectors $Q_{AF}=(H_o,K_o,L_o)=(\pm 0.2+m,\pm 0.6+n,L)/(\pm 0.6+m,\pm 0.2+n,L)$ with $L_o=\pm 1,\pm 3,\cdots$ rlu.
For an energy transfer of $E=10\pm 2$ meV (above the anisotropy
gap of $E=8$ meV, see supplementary information),
spin waves are peaked at the expected eight AF Bragg
positions $Q_{AF}$ around $Q=(0,0,\pm 1)$ rlu as shown in Fig. 2a.  Upon increasing energies to
$E=26\pm 2$ (Fig. 2b) and $30\pm 2$ meV (Fig. 2c), spin waves from the two chiralities centered
around the $Q_{AF}$ positions become apparent and increase in size with increasing energy.  The two spin wave rings from
the left and right AF chiralities (Figs. 1a-1d)
 meet near $E=45\pm 3$ meV (Fig. 2d).  At $E=55\pm 3$ meV, the overlapping spin waves from both AF chiralities
still form rings around the $Q_{AF}$ positions (Fig. 2e).  Spin waves have evolved into broad rings centered
around $(H_o,K_o,L_o)=(\pm m, \pm n, L_o)$ at $E=70\pm 3$ meV as shown in Fig. 2f, just
before disappearing into the $75\le E\le 95$ meV spin gap (Fig. 1f).  Upon re-emerging from the spin gap at an energy transfer of
$110\pm 10$ meV, spin waves form transversely elongated ellipses centered at the wave vectors $Q=(\pm 1,0)/(0,\pm 1)$ (Fig. 2g), identical to the
AF ordering wave vector of (Ba,Ca,Sr)Fe$_2$As$_2$ \cite{lharriger,jzhao,raewings}.  Finally, at $E=200\pm 20$ meV, an energy well above the
$150\le E\le 170$ meV spin gap, spin waves move into wave vectors $Q=(\pm 1, \pm 1)$ (Fig. 2h), almost identical to the zone
boundary spin waves for BaFe$_2$As$_2$ \cite{lharriger} and Fe$_{1.05}$Te \cite{lipscombe}.

We use a local moment Heisenberg Hamiltonian with the effective nearest (NN or $J_1$, $J_1^\prime$),
next nearest (NNN or $J_2$, $J_2^\prime$), and next next nearest neighbor (NNNN or $J_3$,$J_3^\prime$) magnetic
exchange couplings (Fig. 1a) to fit the observed spin-wave spectra \cite{cao,yizhuang,chenfang,ludai,yu}.
To account for the $\sim$8 meV low-energy spin gap, we add a spin anisotropy term $J_s$ to align spins along the $c$-axis
(see supplementary information). There are 8 spins in each magnetic unit cell (Figs. 1a and 1b),
therefore we should have four spin wave bands in the Brillouin zone.
From Figs. 1 and 2, we see that spin waves exist in three separate energy ranges: the lowest branch starts from $\sim$9 meV to $\sim$70 meV, second from $\sim$80 meV to $\sim$140 meV, and the third branch from $\sim$180 meV to $\sim$230 meV.
The high quality of the spin-wave data allows us to place quantitative constraints on
effective exchange couplings in the Heisenberg Hamiltonian (see supplementary information).
While the low-energy spin waves between $\sim$9 meV to $\sim$70 meV
are acoustic mode arising mostly from AF interactions of the FM blocked spins, the two other branches of excitations are optical
spin waves associated with exchange interactions of iron spins within the FM blocks \cite{yizhuang,chenfang,ludai,yu}.
We have attempted, but failed, to fit
the entire spin wave spectra using only the effective NN and NNN exchange coupling Heisenberg Hamiltonian (see Fig. 3 and supplementary information). For spin-wave fits that include the NNNN exchange coupling $J_3$,
we find that the low energy spin wave band (acoustic band) depends mainly on $J_1^\prime$,$J_2^\prime$, $J_3$, and $J_c$ (the effective $c$-axis exchange coupling), but not $J_1$ and $J_2$.
The second band depends on the $J_2$ heavily and the top band is mainly determined by $J_1$.

For simplicity, we consider each FM block with 4 aligned spins as a net spin $S_{eff}$.  They interact with each other antiferromagnetically (via $J_{eff}$) to form a cuprates-like AF spin structure. There is one spin-wave band for this effective block-spin Heisenberg model, which has an analytical form for spin-wave dispersion (see supplementary information). By comparing the $J_{eff}$ Heisenberg Hamiltonian with those of the
$J_1$-$J_1^\prime$-$J_2$-$J_2^\prime$-$J_3$-$J_3^\prime$ model, we find that spin waves in the first band can be approximately described by the $J_{eff}$ Heisenberg Hamiltonian, where $J_{eff}S_{eff}=(J_1^\prime+2J_2^\prime+2J_3)S/4$ is $\sim$17 meV. This suggests that the low energy band is mainly determined by $J_1^\prime$,$J_2^\prime$, $J_3$, and $J_c$.
 Physically, the lowest energy band corresponds to the block spin waves where the 4 spins fluctuate in phase  and resemble a single spin. Only at high energies, the relative motions within the blocks can be excited, which correspond to the two high energy optical modes. Thus the high energy bands are basically determined by the intra-block couplings $J_1$ and $J_2$.

To quantitatively determine the spin-wave dispersion,
we determined the measured dispersion from a series of high symmetry scans through the $(H_o,H_o, L_o)$ and $(H_o, 1/2H_o-1/2, L)$ directions, where $L_o$ was integrated to improve counting statistics.
  Figures 3a-3c summarize the dispersion of spin waves along the marked directions
on the right panels. For the low-energy acoustic mode, we find a spin anisotropy gap below 8 meV and
counter propagating spin waves for energies above 30 meV (Fig. 3c).
The two high-energy optical spin-wave modes are essentially dispersionless.  The blue and pink solid lines show Heisenberg Hamiltonian fits
to the dispersion curves with and without $J_3$.  The final fitted effective magnetic exchange couplings for spin-wave dispersions are $SJ_1=-36\pm 2$, $SJ_1^\prime=15\pm8$, $SJ_2=12\pm 2$, $SJ_2^\prime=16\pm 5$, $SJ_3=9\pm 5$, $J_3^\prime=0$, $SJ_c=1.4\pm 0.2$, and $SJ_s=0.44\pm 0.1$ meV (see supplementary information for fits with other parameters).  Figure 3d shows energy dependence of the observed local susceptibility \cite{lester} and our calculation using the fitted parameters.  We see that the calculated local susceptibility agrees quite well with the data.
To further compare the data in Fig. 2 with calculated spin waves using fitted effective exchange couplings, we show
in Figure 4 the two-dimensional spin-wave projections in the $[H_o,K_o]$ plane convoluted with instrumental resolution.
The calculated spin-wave spectra capture all essential features in the data.

For a Heisenberg model with spin $S$, the total moment sum rule stipulates $M_0=(g\mu_B)^2S(S+1)$.
For irons in the $3d^6$ electronic state, the maximum possible moment is $gS=4$ $\mu_B$/Fe for $g=2$, giving $M_0=24$ $\mu_B^2$/Fe.
Based on absolute spin wave intensity
measurements in Fig. 3d, the sum of the fluctuating moments below $\sim$250 meV is $\left\langle m^2\right\rangle\approx 16\pm3$ $\mu_B^2$/Fe.
If we assume that the ordered moment is on the order of $\sim$3 $\mu_B$/Fe \cite{wbao1,pomjakushin1,wbao2}, we see that the total moment sum rule is exhausted for magnetic scattering at energies below 250 meV.  Therefore, spin waves in insulating
Rb$_{0.76}$Fe$_{1.63}$Se$_2$ can be regarded as a classic local moment system where a Heisenberg Hamiltonian is an appropriate description
of spin-wave spectra.

It is instructive to compare the effective magnetic exchange couplings in different iron-based superconductors. First,  comparing  Rb$_{0.89}$Fe$_{1.58}$Se$_2$ with  Fe$_{1.05}$Te \cite{lipscombe}, we note that although their static AF orders have
completely different structures, these two iron chalcogenides are very similar in terms of the  values of their effective exchange couplings.
Both of them have: (i) large FM $J_{1}$ (or $J_{1a}$), (ii) large anisotropy between the two NN couplings $J_{1} $($J_{1a}$) and $J_{1}^\prime$ (or $J_{1b}$), (iii) AF NNN couplings and small anisotropy between two NNN couplings $J_{2}$(or, $J_{2a}$) and $J_{2}^\prime$ (or $J_{2b}$), and  (iv) significant AF NNNN couplings $J_3$.  Therefore, the presence of the iron vacancy ordering in Rb$_{0.89}$Fe$_{1.58}$Se$_2$ reduces  magnetic frustration and stabilizes the blocked AF structure, but does not change the local magnetic exchange couplings strengths as compared to Fe$_{1.05}$Te \cite{lipscombe}.  Second, comparing  iron-chalcogenides to iron-pnictides, we find  that there are important differences as well as essential common features:  the differences include the large  FM $J_{1a}$ and significant AF $J_3$ in iron-chalcogenides  against the large AF $J_{1a}$ and negligible $J_3$ in iron-pnictides, respectively,    and  the common features include  the large anisotropy of NN exchange couplings and similar AF NNN couplings.  While the NN exchange couplings vary significantly according to the spin configurations between the corresponding two NN sites in the magnetically ordered states, the AF NNN exchange coupling remains almost uniform amongst different classes of materials even though their AF structures can be quite different.
This is consistent with the idea that $J_2$ is mainly determined by a local superexchange mechanism mediated by As or Se/Te \cite{si2008}.  Regarding the microscopic origin of superconductivity,  the difference between the NN exchange couplings of the two classes of materials suggests that the NN FM  exchange coupling cannot be responsible for superconductivity since electron pairing is in the spin singlet channel \cite{yu2011}, which is not allowed by the FM coupling. However, the similarity on  $J_2$  in both classes of materials suggests that if superconductivity in all Fe-based materials has a common magnetic origin,   it must be intimately associated with the
 NNN magnetic exchange interactions, likely resulting in a $s$-wave pairing symmetry \cite{chen2011}.

We thank Masaaki Matsuda for his help on triple-axis measurements discussed in the supplementary information.
The neutron scattering work at UT is supported
by the U.S. NSF-OISE-0968226, and by the U.S. DOE, Division of Scientific User Facilities (P.D.).
The single crystal growth effort at UT is supported by U.S. DOE BES under Grant No. DE-FG02-05ER46202 (P.D.).
Work at IOP is supported by the Chinese Academy of Sciences. D.X.Y. is supported by NSFC-11074310.


\onecolumngrid
\newpage
\title{Supplementary Information: Spin Waves and magnetic exchange interactions in insulating Rb$_{0.89}$Fe$_{1.58}$Se$_2$}
\author{Miaoyin Wang}
\author{Chen Fang}
\author{Dao-Xin Yao}
\author{GuoTai Tan}
\author{Leland W. Harriger}
\author{Yu Song}
\author{Tucker Netherton}
\author{Chenglin Zhang}
\author{Meng Wang}
\author{Matthew B. Stone}
\author{Wei Tian}
\author{Jiangping Hu}
\author{Pengcheng Dai}


\maketitle
\section{Supplementary data}

In addition to the spin wave data presented in the main text, we have taken triple-axis spectrometer measurements on HB-1 at Oak Ridge National Laboratory
to determine the low-energy spin anisotropy gap. Before showing the results, we note that although the scattering cross section is related to the dynamic structure factor $S(Q,E)$, it is proportional to the imaginary part of the dynamic susceptibility $\chi^{\prime\prime}(Q,\omega)$ if the temperature is much lower than the lowest energy spin waves. Theoretically, one has $S(Q,E)= 1/(1-\exp(-E/(k_BT)))\chi^{\prime\prime}(Q,E)$. If $k_BT<<E$ as is the case of the experiment, one has $S(Q,E)\propto\chi^{\prime\prime}(Q,E)$. Figure 5(a) shows $\chi^{\prime\prime}(Q,E)$ at $Q_{AF}=(0.6,0.2,3)$, which clearly establishes the anisotropy spin gap of $\sim$8 meV.
Constant energy scans at 5 meV and 10 meV shown in Fig. 5(b) confirm the presence of the spin gap below 8 meV.
To further demonstrate the presence of spin gaps around 80 meV and 160 meV,
we show in Figs. 5(c)-(e) constant energy cuts for energies of $E=74\pm4$, $82\pm 4$, and $90\pm 4$ meV, respectively.
There are clearly no magnetic scattering near $E=82\pm 4$ meV [Fig. 5(d)].  Figures 5(f)-(h) show similar constant-energy
images at $E=140\pm 10$, $155\pm 15$, and $195\pm 15$ meV.  The scattering near $E=155\pm 15$ meV are featureless, confirming the presence of a spin gap at this energy.

\section{Model Heisenberg Hamiltonian}

The model we use to understand the magnetic excitation is a quantum spin model with up to third nearest neighbor (NNNN) exchange in the $ab$-plane, nearest neighbor (NN) exchange along the $c$-axis and a single ion anisotropy term, i.e.,\bea H=H_{ab}+H_c+H_s,\eea where \bea H_c&=&J_c\sum_r\mathbf{S}_r\cdot\mathbf{S}_{r+z},\\
\nonumber H_s&=&\frac{J_s}{2}\sum_r(S_{r,x}^2+S_{r,y}^2),\eea and $H_{ab}$ is given in Ref. \cite{chenfang}.
To solve the Hamiltonian, one can use standard linear spin wave approach. A generic position of the spin is given by \bea \mathbf{r}=m\mathbf{l_1}+n\mathbf{l_2}+\mathbf{d_i},\eea where $m,n$ are integers and \bea\mathbf{l_1}=(2\mathbf{x}-\mathbf{y})/\sqrt{5},\\
\nonumber\mathbf{l_2}=(\mathbf{x}+2\mathbf{y})/\sqrt{5},\\
\nonumber\mathbf{d_1}=0,\;\mathbf{d_2}=\mathbf{x},\;\mathbf{d_3}=\mathbf{x}+\mathbf{y},\;\mathbf{d_4}=\mathbf{y}.\eea
The Holstein-Primakoff transform (truncated) of the spin operators is given by\\
For $m+n=$even:\bea S_+(\mathbf{r})&=&\sqrt{2S}a_i(\mathbf{R}),\\
\nonumber S_-(\mathbf{r})&=&\sqrt{2S}a^\dag_i(\mathbf{R}),\\
\nonumber S_z(\mathbf{r})&=&S-a^\dag_i(\mathbf{R})a_i(\mathbf{R});\eea
For $m+n=$odd:\bea S_+(\mathbf{r})&=&\sqrt{2S}a^\dag_i(\mathbf{R}),\\
\nonumber S_-(\mathbf{r})&=&\sqrt{2S}a_i(\mathbf{R}),\\
\nonumber S_z(\mathbf{r})&=&-S+a^\dag_i(\mathbf{R})a_i(\mathbf{R}).\eea Define $\psi^\dag(k)=(a^\dag_1(k),a^\dag_2(k),a^\dag_3(k),a^\dag_4(k),a_1(-k),a_2(-k),a_3(-k),a_4(-k))$, and we have\bea H=\frac{1}{2}\sum_k\psi^\dag(k)\left(
                                 \begin{array}{cc}
                                   A(k) & B(k) \\
                                   B(k) & A(k) \\
                                 \end{array}
                               \right)
\psi(k).\eea $A(k)$ and $B(k)$ are four-by-four matrices, defined by:\bea A(k)=S\left(
                                                                                   \begin{array}{cccc}
                                                                                     E_0 & J_1e^{ik_x} & J_2e^{ik_x+ik_y}+J'_3e^{-i2k_x} & J_1e^{ik_y}\\
                                                                                      .& E_0 & J_1e^{ik_y} & J_2^{-ik_x+ik_y}+J'_3e^{-2ik_y} \\
                                                                                      .& . & E_0 & J_1e^{-ik_x}\\
                                                                                      .& . & . & E_0 \\
                                                                                   \end{array}
                                                                                 \right),\eea
\bea B(k)=S\left(
                                                                                   \begin{array}{cccc}
                                                                                       2J_c\cos(k_z)& J'_2e^{-ik_x+ik_y}+J_3e^{-2ik_y} & J'_1e^{-ik_y} & J'_2e^{-ik_x-ik_y}+J_3e^{2ik_x}\\
                                                                                      .& 2J_c\cos(k_z) & J'_2e^{-ik_x-ik_y}+J_3e^{2ik_x} & J'_1e^{ik_x} \\
                                                                                      .& . & 2J_c\cos(k_z) & J'_2e^{ik_x-ik_y}+J_3e^{2ik_y}\\
                                                                                      .& . & . & 2J_c\cos(k_z) \\
                                                                                   \end{array}
                                                                                 \right),\eea
where $E_0=-(2J_1+J_2-J'_1-2J'_2-2J_3+J'_3-2J_c-J_s)S$. The lower triangle elements are suppressed because both matrices are hermitian.

\begin{figure}[t]
\includegraphics[scale=.6]{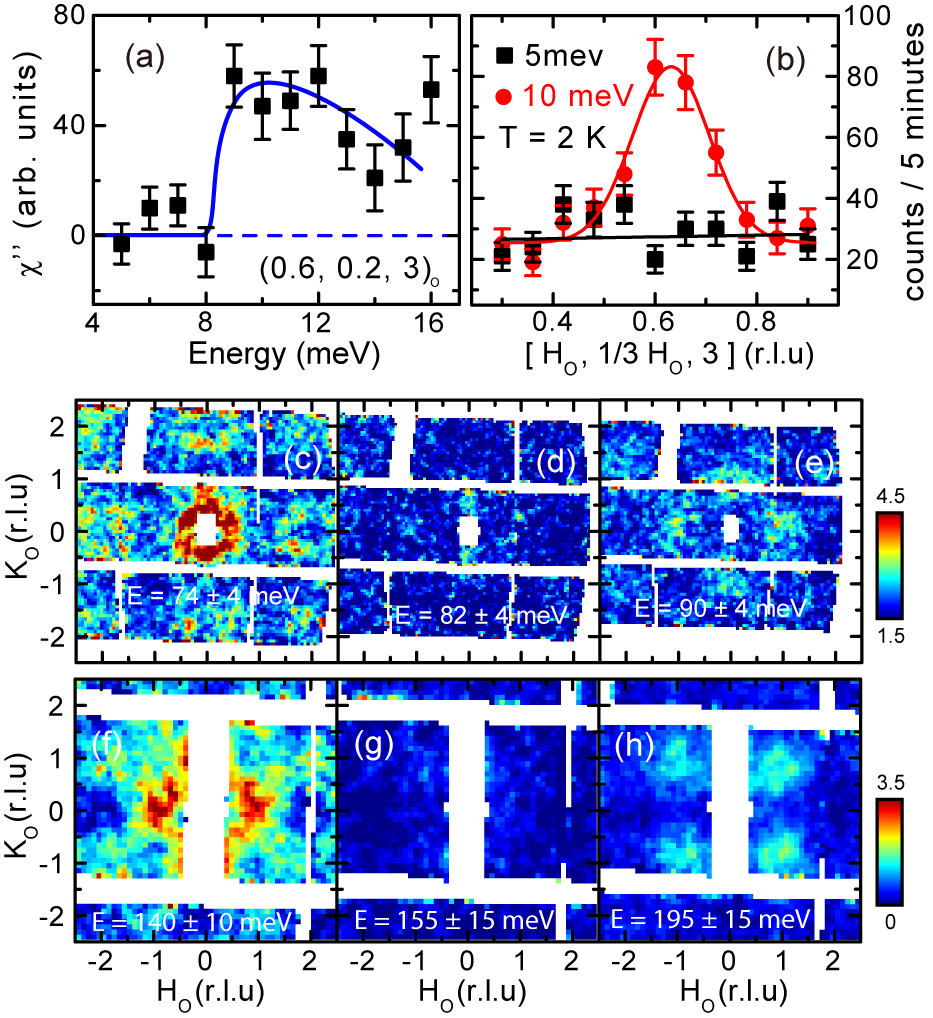}
\caption{Triple-axis spectrometer data obtained on HB-1 and additional spin-wave images near spin gaps on Rb$_{0.89}$Fe$_{1.58}$Se$_2$.
(a) Constant-$Q$ scan at the AF wave vector $Q_{AF}=(0.6,0.2,3)$ rlu with background subtracted, and corrected for Bose population factor.
There is a clear spin gap below $E=8$ meV.  (b) Constant-energy scans across the AF wave vector
at $E=5$ meV and $E=10$ meV.  The data confirm the presence of a spin gap at 5 meV.
Spin wave images in the $(H_o,K_o)$ plane for energy transfers of (c) $E=74\pm 4$; (d) $82\pm 4$;(e) $90\pm 4$;
(f) $140\pm 10$; (g) $155\pm 15$; $195\pm 15$ meV.  There are clearly no spin wave excitations at $E=82\pm 4$ and $155\pm 15$ meV.
}
\end{figure}

\begin{figure}[t]
\includegraphics[scale=.48]{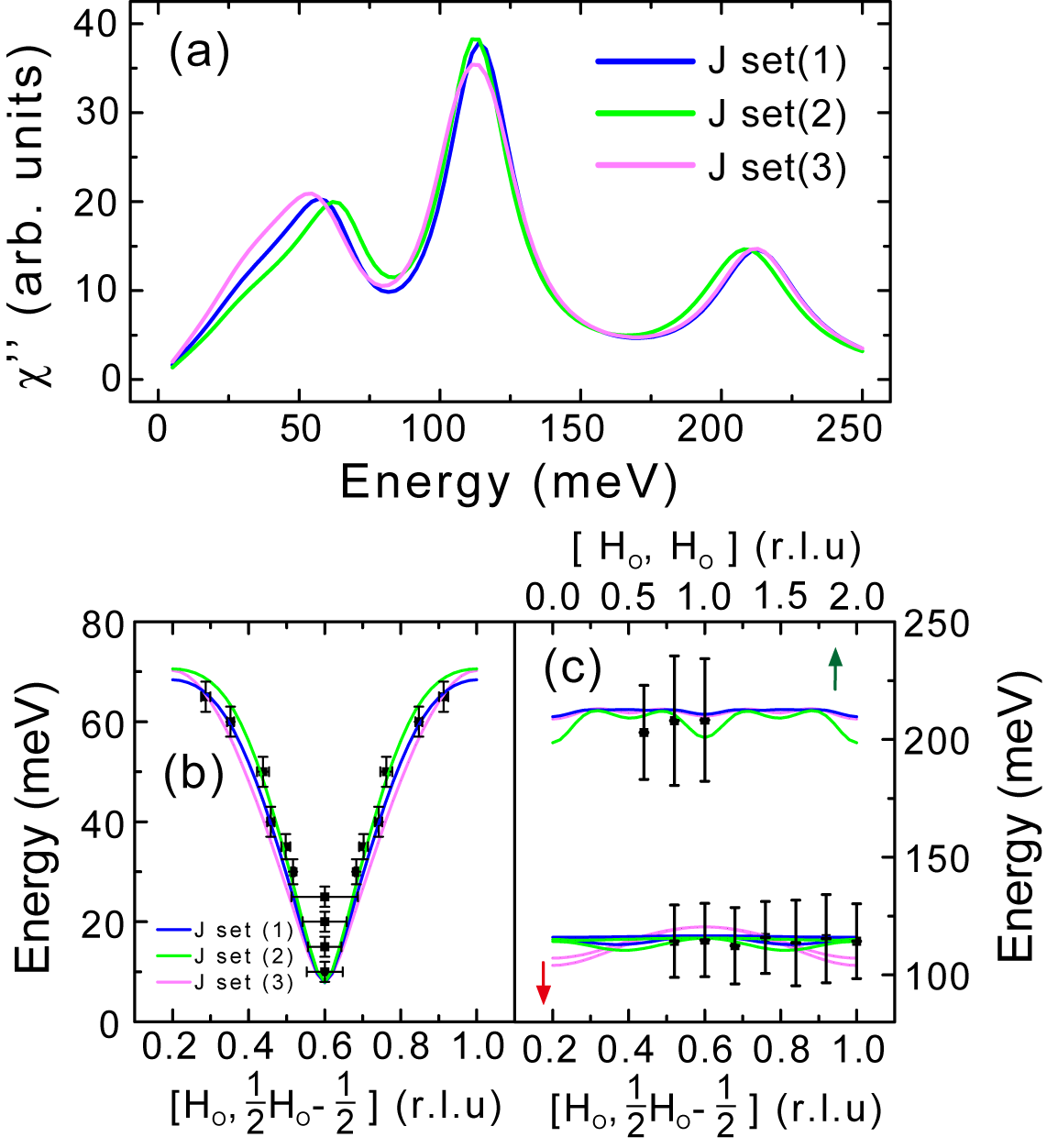}
\caption{(a) Energy dependence of imaginary part of local susceptibility for the three different exchange parameter sets. (b,c) Dispersion curves for
the three different exchange parameter sets as discussed in the text.
}
\end{figure}

We use equations of motion to solve this Hamiltonian.\bea\partial{\psi(k)}/\partial{t}=-i\left(
                                                                                           \begin{array}{cc}
                                                                                             A(k) & B(k) \\
                                                                                             -B(k) & -A(k) \\
                                                                                           \end{array}
                                                                                         \right)\psi(k).\eea
Solving this eigenvalue problem for each $k$, we have\bea H=\sum_{i=1,2,3,4;k}(\gamma^\dag_i(k)\gamma_i(k)+1/2)\omega_i(k),\eea and\bea a_i(k)=\sum_j U_{ij}(k)\gamma_j(k)+V_{ij}(k)\gamma^\dag_j(-k).\eea The differential cross section of inelastic neutron scattering can be expressed in terms of the spin wave dispersion and wave functions:\bea\sigma(\omega,q)=I_0(\omega,q)(1+n_B(\omega,T))\sum_\alpha|\sum_iU_{i\alpha}(q)+V^\star_{i\alpha}(-q)|^2 D(\omega,\omega_\alpha).\eea In the above expression, $I_0(\omega,q)$ includes all factors of experimental resolution extracted from information of each detector, $n_B(\omega,T)$ is the Bose factor and $D(\omega,\omega_\alpha)$ is the harmonic oscillator damping given by \bea D(\omega,\omega_0)=\frac{4}{\pi}\frac{\omega\omega_0\Gamma(\omega)}{(\omega^2-\omega_0^2)^2+4\Gamma(\omega)^2\omega^2}.\eea The damping strength $\Gamma(\omega)$ is approximated by a linear function of energy whose explicit form is to be fitted. Our fitting is based on so far the most general spin model with all symmetry allowed exchanges up to NNNN. A failure of this model in understanding the data would mean that the observed excitations cannot be explained by a local moment picture and the effect of itinerant electrons must be seriously considered.

\begin{figure}[t]
\includegraphics[scale=.48]{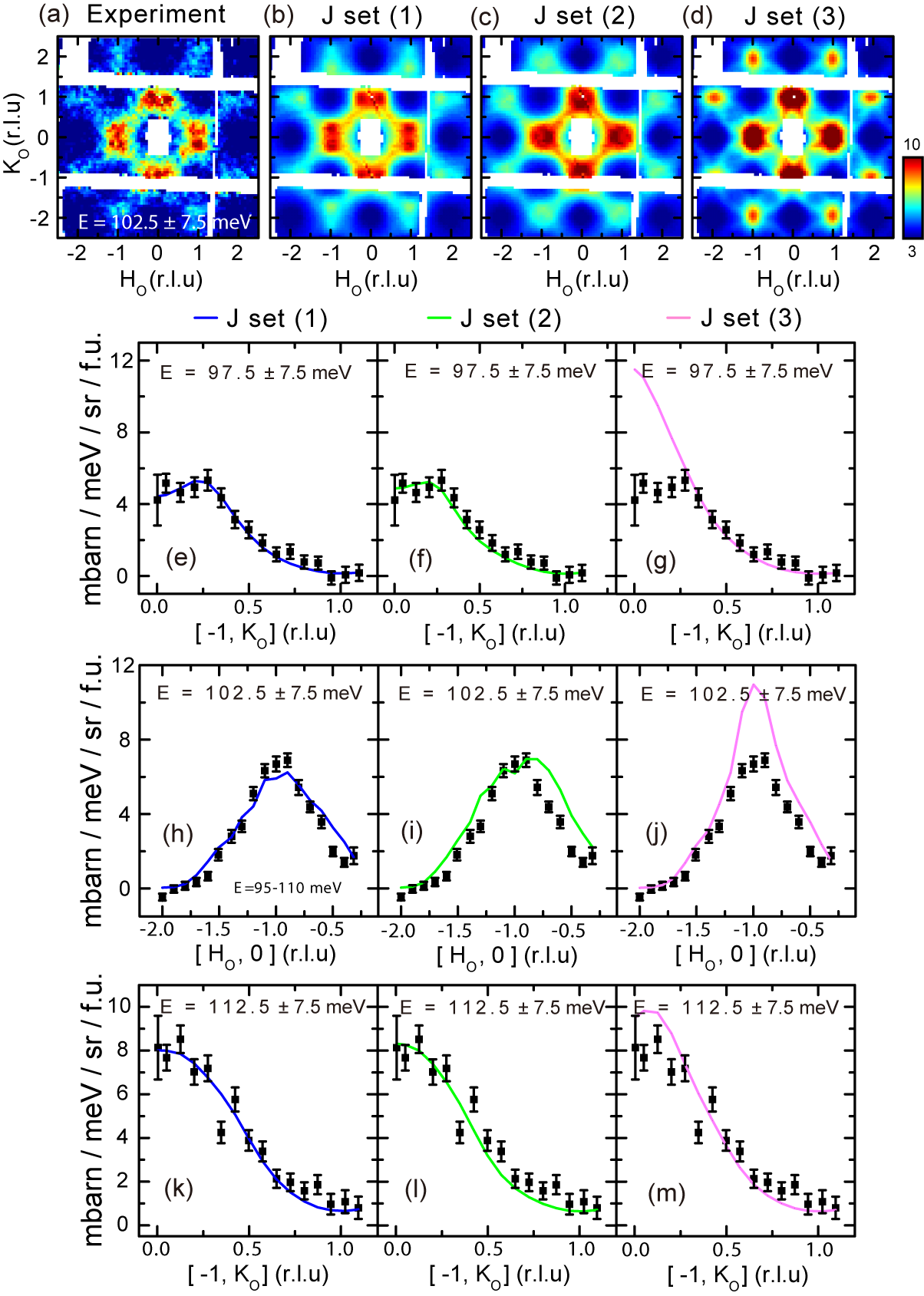}
\caption{Calculated dynamic structure factor and their comparison with Heisenberg Hamiltonian with different exchange parameters.
(a) Constant energy cut of data at $E=102.5\pm 7/5$ meV projected onto the $(H_o,K_o)$ plane. (b,c,d) Calculated dynamic structure factor $S(q,\omega=102.5\pm7/5)$ projected
onto the  $(H_o,K_o)$ plane for three different exchange coupling parameters. (e-m)Cuts along different directions and their comparison with
spin wave calculations in three different exchange coupling parameters.
}
\end{figure}

\section{Fitting constraints}
The high quality of the data allows one to place quantitative constraints on parameters in the model. The data shows that the excitations exist in three separate energy ranges. The lowest branch starts from $\sim9$ meV to $\sim70$ meV, second from $\sim100$ meV to $\sim140$ meV and the third branch from $\sim$180 meV to $\sim$230 meV. The low energy part of the first branch can be fitted very well by the form\bea \epsilon(k)=\sqrt{{\Delta^{exp}_s}^2+{v^{exp}_s}^2k^2},\eea with $v^{exp}_s=300\ {\rm meV\cdot\AA}$ and $\Delta^{exp}_s=9$ meV. At the propagation vector of the ground state $Q=(0.6,0.2,1)$ rlu (in the orthorhombic basis), energy has $k_z$ dispersion, and the band top is about $E_c^{exp}\sim$30 meV. All these values have analytical expressions in the spin wave model. The anisotropy gap (bottom of the first branch) is\bea\Delta_s=S\sqrt{J_s(2J'_1+4J'_2+4J_3+4J_c+J_s)}.\eea The top of the first band is reached at $Q_o=(0.2,0.4,0)$ rlu with \bea E_{1t}&=&2S[2J_1^2+J'_1(J'_2+J_3-J'_3+J_c)+(J'_2+J_3-J'_3)(J'_2+J_3-J'_3+2J_c)-J_1(J'_1+2(J'_2+J_3-J'_3+J_c))\\
\nonumber&&-\sqrt{4J_1^4+{J'_1}^2J_c^2+4J_1^2(J'_2+J_3-J'_3+J_c)(J'_1+J'_2+J_3-J'_3+J_c)-4J_1^3(J'_1+2(J'_2+J_3-J'_3+J_c))}]^{\frac{1}{2}}.\eea
Without single ion anisotropy, i.e., $J_s=0$, the spin wave velocity is given by\bea
v_s=\sqrt{\frac{5}{2}}S\{[J_1(J'_1+2(J'_2+J_3-J'_3))+J'_1(J_2-J'_2-J_3+J'_3)+2(J_2(J'_2+J_3-J'_3)+2J_3(J'_3-J'_2))]\\
\nonumber(J'_1+2(J'_2+J_3+J_c))/(J_1-J'_1+J_2-J'_2-J_3+J'_3)\}^{1/2}.\eea The expression with $J_s\neq0$ is also available but too lengthy to be placed here, and interested readers can request it from the authors. The second branch actually contains two close spin wave bands. The branch starts at $Q=(0.3,0.1,1)$ rlu with energy $E_{2b}$, whose expression is again too lengthy to be published. The second branch ends at $\Gamma=(0,0,0)$ point with \bea E_{2t}=S\sqrt{(2J_1-2J'_1+2J_2-2J'_2-2J_3+2J'_3-J_s)(2J_1+2J_2-2J'_2-2J_3+2J'_3-4J_c-J_s)}.\eea
The highest branch starts at $\Gamma$ point with \bea E_{3b}=S\sqrt{(4J_1-4J'_2-4J_3-J_s)(4J_1-2J'_1-4J_c-J_s)},\eea and ends at $(0.2,0.4,0)$ with \bea E_{3t}&=&2S[2J_1^2+J'_1(J'_2+J_3-J'_3+J_c)+(J'_2+J_3-J'_3)(J'_2+J_3-J'_3+2J_c)-J_1(J'_1+2(J'_2+J_3-J'_3+J_c))\\
\nonumber&&+\sqrt{4J_1^4+{J'_1}^2J_c^2+4J_1^2(J'_2+J_3-J'_3+J_c)(J'_1+J'_2+J_3-J'_3+J_c)-4J_1^3(J'_1+2(J'_2+J_3-J'_3+J_c))}]^{\frac{1}{2}}.\eea
The band top along the $c$-axis is reached at $(0.6,0.2,0)$ with\bea E_c=S\sqrt{[2(J'_1+2J'_2+2J_3)-J_s](4J_c+J_s)}.\eea Based on the data and considering the effect of large damping at high energies, we have for the above quantities the following constraints:\bea
\Delta_s&=&\Delta_s^{exp}=8\sim12\ {\rm meV},\\
\nonumber v_s&=&v_s^{exp}=250\sim300\ {\rm meV\cdot\AA},\\
\nonumber E_{1t}&=&60\sim75\ {\rm meV},\\
\nonumber E_{2b}&=&90\sim110\ {\rm meV},\\
\nonumber E_{2t}&=&110\sim130\ {\rm meV},\\
\nonumber E_{3b}&=&180\sim200\ {\rm meV},\\
\nonumber E_{3t}&=&200\sim220\ {\rm meV},\\
\nonumber E_c&=&25\sim30\ {\rm meV}.\eea

\section{Fitting parameters}
The above constraints give a very narrow range of parameters, we can further constraint
possible exchange constants so that
a quantitative fit to the data shown in the paper can be found. In this section we discuss what elements are indispensable to our fittings.

\begin{figure}[t]
\includegraphics[scale=.48]{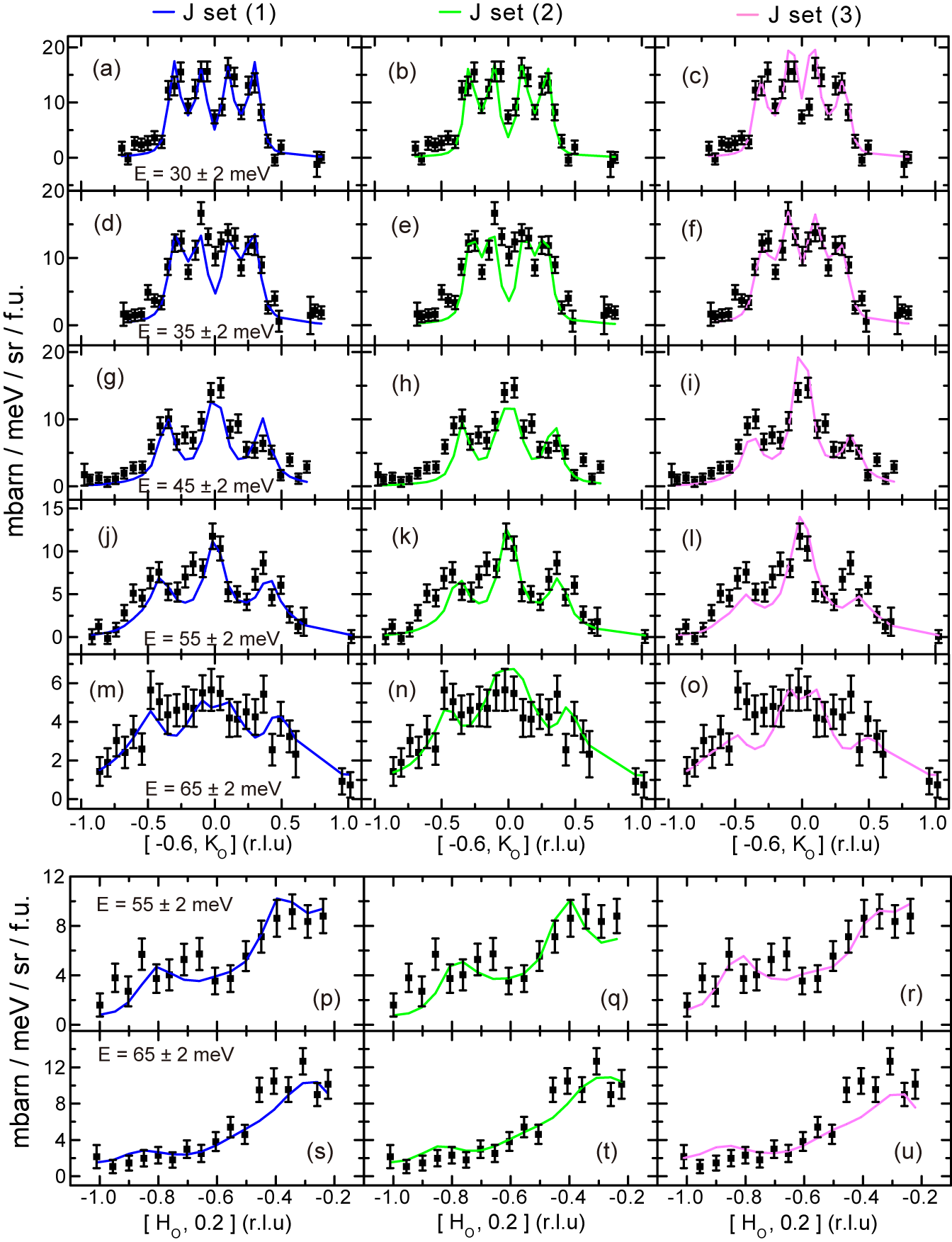}
\caption{ Spin-wave dispersions of Rb$_{0.89}$Fe$_{1.58}$Se$_2$ and fits using Heisenberg Hamiltonian with
three different exchange coupling parameters as discussed in the text.
Spin-wave dispersions in the acoustic branch obtained by cutting along high-symmetry directions
and model fits using three different sets of exchange coupling parameters (a)-(o)
Cuts along the $[-0.6, K_o]$ direction by integrating $H_o$ from $-0.65$ to $-0.55$.
(p)-(u) Cut along $[H_o, 0.2]$ direction by integrating $K_o$ from 0.15 to 0.25.
}
\end{figure}

We first emphasize that a proper fitting should have $J_3>0$ and $J'_1>0$ (antiferromagnetic).
To see this, we compare the following possible parameters since they can all approximately describe the data:

(1)	$SJ_1 = -36$, $SJ_1^\prime = 15$, $SJ_2 = 12$, $SJ_2^\prime = 16$, $SJ_3 = 9.5$, $SJ_3^\prime = 0$, $SJ_c = 1.4$,
$SJ_s = 0.44$ meV.
(2)	$SJ_1 = -36$, $SJ_1^\prime = -5.7$, $SJ_2 = 13.4$, $SJ_2^\prime = 22.4$, $SJ_3 = 14.2$, $SJ_3^\prime = 0$, $SJ_c = 1.4$,
$SJ_s = 0.44$ meV.
(3)	$SJ_1 = -36$, $SJ_1^\prime = 10$, $SJ_2 = 11$, $SJ_2^\prime = 28.7$, $SJ_3 = 0$, $SJ_3^\prime = 0$, $SJ_c = 1.4$,
$SJ_s = 0.44$ meV.
Figure 6 summarizes the calculated $\chi^{\prime\prime}(\omega)$ and spin wave dispersions for all three sets of parameters.  From the calculation, we see that all three parameter sets give similar local susceptibilities, and therefore cannot be distinguished based on $\chi^{\prime\prime}(\omega)$ alone.

By comparing the calculated spin wave dispersion curves with data, we were able to separate which model is correct.  Figure 6(b) and (c)
shows the outcome for the three sets of exchange couplings for the acoustic and optical modes, respectively.  We see that parameters of (1) and (2) fit the acoustic and optical data slightly better.
Although the imaginary part of local susceptibility and dispersion curves for different exchange parameter sets are similar,
their constant energy patterns at $\sim$110 meV are very different, which provides key clues to the choice among different exchange coupling
parameters. In the energy range around 110 meV, several optical branches are mixed together.
The combined spin wave intensity patterns depend sensitively on the exchange coupling parameters. Figure 7 compares directly the calculated patterns with the observation for the three set of exchange parameters.  Clearly, the first set of parameters describes the data much better.  This is what we have used to determine the effective magnetic exchange coupling constants. This conclusion is further confirmed by comparing the calculated dispersion with the observed dispersion using the three sets of parameters as shown in Figs. 7, 8 and 9.

As a remark, we note the important fact that the in-block NNN exchange $J_2$ must be positive (antiferromagnetic) for all candidate sets of parameters. $J_2$ has little effect on the first and the third branches of dispersion, but is strongly coupled to the middle branch. A ferromagnetic $J_2$ can push up the second branch for about 30\%. This means the gap between first and second branches would be more than 40 meV, while in experiment it is clearly less than 30 meV.

\begin{figure}[t]
\includegraphics[scale=.48]{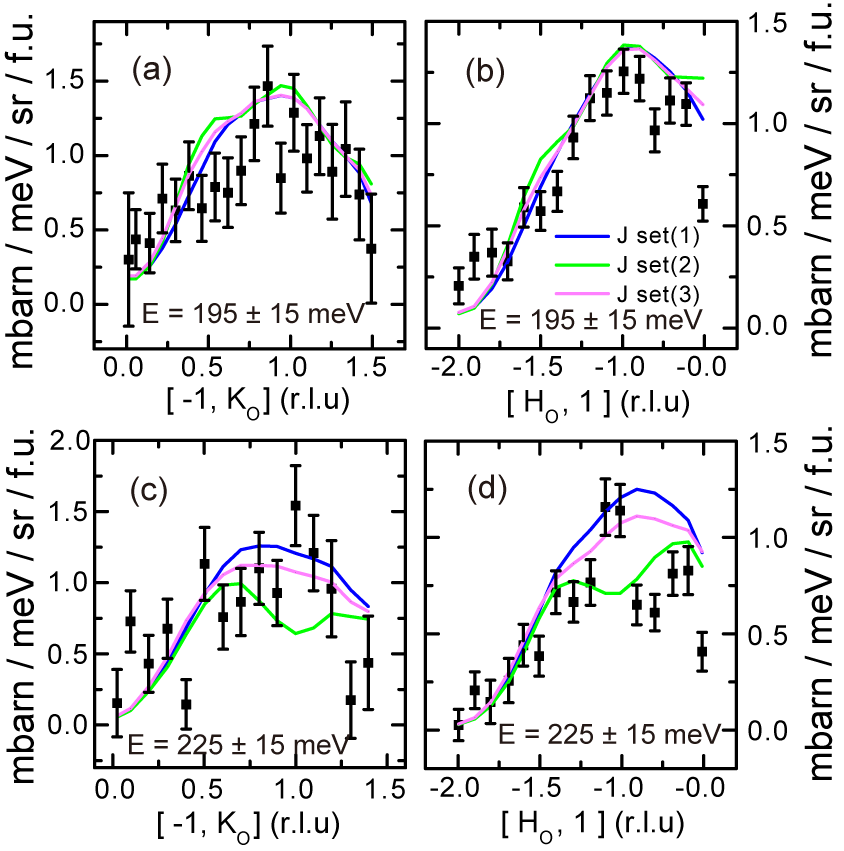}
\caption{ Spin-wave cuts of Rb$_{0.89}$Fe$_{1.58}$Se$_2$ and fits using Heisenberg Hamiltonian with three sets of parameters.
$Q$-cuts for the highest branch. The $H_o$ integration range in (a),(c) is from $-1.05$ to $-0.95$.
Integration of the $K_o$ range in (b),d) is from 0.8 to 1.2.
}
\end{figure}

\section{Sum rule}
Here we discuss the total moment sum rule. For a Heisenberg model
with spin $S$, the sum rule is formulated as Ref. \cite{lorenzana}:
\begin{equation}
M_{o} = \frac{1}{N}\sum_{\alpha} \int
d\mathbf{k}\int_{-\infty}^{\infty} d \omega S^{\alpha \alpha}
(\mathbf{k},\omega) \nonumber \\
=M_x+M_y+M_z\nonumber\\
=g^2\mu_B^2 S(S+1),
\end{equation}
where $g$ is the Lande factor. For free electrons $g=2$.  In Rb$_{0.89}$Fe$_{1.58}$Se$_2$, the maximum possible spin $S=2$ is expected, which gives $M_{o} =24$ $\mu_B^2$/Fe.

The longitudinal part $M_z$ comes from the static moment (elastic)
and the inelastic contribution. For our system, the static moment is
about $3$ $\mu_B$/ Fe \cite{wbao1}, which contributes $9$ $\mu_B^2$/Fe. The
inelastic part mainly comes from the two-magnon scattering process.
The magnetization reduction can be evaluated as $\Delta S=0.5$ from
the static moment for $S=2$.  From Ref. \cite{lorenzana}, we can estimate the two-magnon spectral weight as $\Delta
S(1+\Delta S)g^2\mu_B^2 \simeq 3\mu_B^2$/Fe, where the normalization
factor has been chosen as $1$. The spectral weight from the
two-magnon process is only $1/3$ of the elastic part, which is much
weaker than the cuprates which has $S=1/2$. In unpolarized neutron
experiments, the two-magnon spectral weight is generally very hard
to detect. We will ignore it in the following treatment.

The transverse part $M_x+M_y$ mainly comes from the one-magnon spin
wave spectrum. According to Eq. (1) in Ref. \cite{lester}, we can get the dynamic structure factor
$S(E)$ by removing the magnetic form factor. Then using Eq. (5) in Ref. \cite{lester},
we can get the transverse part by integrating $S(E)$ over
the whole energy range. Experimentally we do not observe the neutron
scattering signal above $250$ meV, so we can choose the integration
range from $8$ to $250$ for the inelastic signal only. We get the
transverse part $\sim 26\pm 5$ $\mu_B^2 (f.u.)^{-1}$, where $f.u.$
means formula unit. Considering the formula of
Rb$_{0.89}$Fe$_{1.58}$Se$_2$, we divide it by a factor of $1.6$. The
transverse part $M_x+M_y$ is evaluated as $16\pm 3\mu_B^2/Fe$.

The total moment from our evaluation is $25\pm 5$  $\mu_B^2$/Fe, which is very close to the expected total moment from the sum rule. Thus the Heisenberg model with $S=2$ is an appropriate description for the insulating Rb$_{0.89}$Fe$_{1.58}$Se$_2$ and the spin waves describe the spin dynamics very well.

\end{document}